\documentclass[aps,showpacs, nofootinbib,floatfix]{revtex4}
\usepackage{graphicx, graphics, color}
\usepackage{subfig} 

\input amssym.tex
\input amssym.def

\newcommand{\be}{\begin{equation}}
\newcommand{\ee}{\end{equation}}
\newcommand{\ben}{\begin{eqnarray}}
\newcommand{\een}{\end{eqnarray}}

\newcommand{\cL}{{\cal L}}

\newcommand{\p}{\partial}
\newcommand{\na}{\nabla}
\newcommand{\tna}{\tilde \nabla}

\newcommand{\tg}{{\tilde g}}
\newcommand{\tq}{{\tilde q}}
\newcommand{\tth}{{\tilde h}}

\newcommand{\tG}{{\tilde G}}
\newcommand{\tR}{{\tilde R}}
\newcommand{\tH}{{\tilde H}}

\newcommand{\trho}{{\tilde \rho}}

\newcommand{\ga}{\gamma}

\begin{document}

\title{Uniqueness of photon sphere for Einstein-Maxwell-dilaton black holes with arbitrary coupling constant}


\author{Marek Rogatko} 
\email{rogat@kft.umcs.lublin.pl,
marek.rogatko@poczta.umcs.lublin.pl }
\affiliation{Institute of Physics \protect \\
Maria Curie-Sklodowska University \protect \\
20-031 Lublin, pl.~Marii Curie-Sklodowskiej 1, Poland}

\date{\today}

\begin{abstract}
The uniqueness of static asymptotically flat photon sphere for staic black hole solution in Einstein-Maxwell-dilaton theory with arbitrary coupling constant was proposed.
Using the conformal positive energy theorem we show that the dilaton photon sphere subject to the non-extremality condition authorizes
a cylinder over a topological sphere.

 \end{abstract}

\maketitle

\section{Introduction}
Gravity theories like general relativity and its modifications predict existence of a spacetime region in which photon orbits are closed.
The aforementioned regions play an essential role in gravitational lensing, one of the main tools in astrophysical observations \cite{iye85}-\cite{vir00}.

The photon sphere can be considered as a timelike hypersurface on which the bending angle is unboundedly large. The compact objects like black holes,
neutron stars, wormholes and others ought to be in principle surrounded by a photon sphere. On the other hand,  as was revealed in Ref.\cite{car14}-\cite{dec10}
the photon spheres are connected with quasi-normal modes for the compact objects in question. Their presences are the main factors for their stabilities.

Moreover, it was found that photon spheres possess some very intriguing features, like the ones of the lapse function constancy on it. They are totally umbilical
hypersurfaces with constant mean curvature and constant surface gravity \cite{cla01}-\cite{ced14}. The properties in question very much resemble the characteristic features of black hole event horizons.
In the case of black hole physics the {\it no-hair} theorem its mathematical formulation, the uniqueness theorem 
resolves the problem of classification of domains of outer communication of suitably regular black hole spacetimes. \\
The first attempts to classify non-singular static black hole solutions in Einstein gravity was undertaken in \cite{isr}
and some others mathematical refinements were presented in Refs.\cite{mil73}-\cite{he93}.
The complete classification of static vacuum and electro-vacuum black hole solutions was finished 
in \cite{chr99a,chr99b}, where the condition of non-degeneracy of the 
event horizon was removed  as well as it was proved that all degenerate components of 
the black hole event horizon have charges of the same signs.
As far as stationary axisymmetric black holes is concerned, the problem turned out to be far more complicated
\cite{stat} and the complete uniqueness proof was achieved by Mazur \cite{maz} and Bunting \cite{bun}
(see for a review of the uniqueness of black hole
solutions story see \cite{book} and references therein).
The contemporary unifications schemes as M/string theories triggered the efforts of classification of higher dimensional 
charged black holes both with non-degenerate and degenerate component 
of the event horizon was proposed in Refs.\cite{gib02}-\cite{rog}. On the other hand, some progress concerning the nontrivial case 
of $n$-dimensional rotating black objects (black holes, black rings or black lenses) uniqueness theorem were 
presented in \cite{nrot}, while the behavior of matter fields in the spacetime of higher dimensional 
black hole was examined in \cite{rog12}.
The desire of constructing a consistent quantum gravity theory raised also interests 
in mathematical aspects of black holes in the low-energy limit of the string theories
and supergravity \cite{sugra1, sugra}. Various modifications of Einstein gravity 
such as the Gauss-Bonnet extension were examined from the point of view of black hole uniqueness theorem.
The strictly stationary static vacuum spacetimes was discussed in \cite{shi13a}, while it turned out that
up to the small curvature limit, static uncharged or electrically charged Gauss-Bonnet black hole is diffeomorphic to 
Schwarzschild-Tangherlini or Reissner-Nordstr\"om black hole solution, respectively \cite{rog14}. 
For black holes appearing in Chern-Simons modified gravity it was proved that a static asymptotically flat black hole
solution was unique to be Schwarzschild spacetime \cite{shi13}, while electrically 
charged black hole in the theory in question was diffeomorphic to
Reissner-Nordstr\"om black hole \cite{rog13}.

The analogy of the photon sphere and the black hole event horizon arises a tantalizing question if the presence of a photon sphere
uniquely characterizes the spacetimes with asymptotical charges. This problem was tackled for the first time in \cite{ced14}, where 
it was shown that an asymptotically flat vacuum Einstein equations with mass and having a photon sphere is isometric to
Schwarzschild solution characterizing by the same mass. Recently, modified version of arguments presented by Bunting and Massod \cite{ced15}
was applied in the proof of the uniqueness of Einstein vacuum photon sphere and electro-vacuum one \cite{ced15a}. On the other hand,
the uniqueness of static Einstein scalar and Einstein-Maxwell spacetimes with a photon spheres were given in Refs.\cite{yaz15a, yaz15b}.
In Ref.\cite{yaz15bb} the general classification of photon spheres (covering black hole and non-black hole spacetimes) in Einstein-Maxwell-dilaton gravity with arbitrary coupling constant
was proposed, subject to the auxiliary condition that the lapse function regulary foliates the spacetime outside the photon sphere. 
In our paper we relax the additional condition and consider the case of
the uniqueness theorem for dilaton black hole photon sphere in the dilaton gravity with arbitrary coupling constant. Namely, we shall pay attention to
the system described by the standard action
\be
I = \int d^4 x \sqrt{-{\hat g}} \bigg[ {}^{(4)}R - e^{-2 \alpha \phi} F_{\mu \nu}F^{\mu \nu} - 2~\na_\mu \phi \na^\mu \phi  \bigg],
\label{act}
\ee
where ${\hat g}_{ij}$ stands for four-dimensional metric tensor, $F_{\mu \nu}$ is the strength tensor of $U(1)$-gauge Maxwell field, $\phi$ is the dillaton field and 
$\alpha$ is the coupling constant.
Because of the fact that our considerations will be bounded with the static spacetime,  let us suppose that there exists a smooth Riemannian manifold and a smooth lapse function
$N \rightarrow M^3 \rightarrow R^+$, such that $M^4 = R \times M^3$. The line element of the above manifold is provided by the following:
\be
ds^2 = {\hat g}_{\alpha \beta} ~dx^\alpha dx^\beta = - N^2~dt^2 + g_{ij} dx^i dx^j.
\label{lel}
\ee 
Moreover, we introduce asymptotically timelike Killing vector field 
$k_{\alpha} = \big({\p \over \p t }\big)_{\alpha}$, such that the $U(1)$-gauge Maxwell and the dilaton field are invariant under the action generated by this Killing vector field, i.e.,
$\cL_{k} F_{\alpha \beta} = 0, ~\cL_{k} \phi =0$.
The above conditions define the $U(1)$-gauge field and dilaton field staticity. The above notions of staticities, i.e., metric staticity and field staticity, are consistent.
It follows from the fact that one has the Ricci-static spacetime which means that the Ricci one-form is proportional to the Kiling vector field $k_\alpha$, which directly follows from
the equations of motion and the field staticity. On the other hand, it can be proved that the static spacetime is Ricci-static \cite{book}.
We shall assume further, that the three-dimensional submanifold $(M^3,~g_{ij})$  is simply connected. This fact enables one to define electric field having potential $\psi$ in the standard
form $E_{\beta} = - F_{\beta \gamma}  k^\gamma = \na_\beta \psi$.
For the readers convenience we also quote the equations of motion for the system in question
\ben \label{eqm1}
{}^{(g)}\na_{i} {}^{(g)}\na^{i} N &=& 
\frac{e^{-2 \alpha \phi}}{ N} {}^{(g)}\na_{i}\psi {}^{(g)}\na^{i} \psi, \\ 
N~{}^{(g)}\na_{i} {}^{(g)}\na^{i} \psi &=&  {}^{(g)}\na_{i} \psi {}^{(g)}\na^{i} N + 2~\alpha ~N~{}^{(g)} \na_{k} \psi {}^{(g)}\na^{k} \phi, \\ \label{m1}
N~{}^{(g)}\na_{i} N {}^{(g)}\na^{i} \phi &+&  N^2~{}^{(g)}\na_{m}{}^{(g)}\na^{m} \phi - \alpha~e^{-2 \alpha \phi}~{}^{(g)}\na_{c} \psi {}^{(g)}\na^{c} \psi = 0,\\ 
{}^{(g)}R_{ij} - \frac{1}{ N}{}^{(g)}\na_{i} {}^{(g)}\na_{j} N &=& \frac{e^{-2 \alpha \phi}}{N^2}
\bigg( g_{ij} {}^{(g)}\na_{m} \psi {}^{(g)}\na^{m} \psi  - 2~{}^{(g)}\na_{i}\psi {}^{(g)}\na_{j} \psi \bigg) 
+ 2~{}^{(g)}\na_{i}\phi {}^{(g)}\na_{j} \phi, 
\een
where the covariant derivative with respect to the metric tensor $g_{ij}$ is denoted by ${}^{(g)}\na$,
while ${}^{(g)}R_{ij}$ is the Ricci tensor defined in $M^3$ space.

Our paper is organized as follows. In Sec.II, after describing the basic features of {\it photon sphere} in Einstein-Maxwell-dilaton gravity
with arbitrary coupling constant, we conduct the uniqueness proof using the conformal positive energy theorem.
Sec.III concludes our investigations.
\section{Uniqueness}
Before we proceed to the main subject of our work, let recall some basic fact which will be useful in our construction of the proof.
First of all, let us assume that the space time under consideration will be asymptotically flat, which means that the spacetime  
contains a data set $(\Sigma_{end}, g_{ij}, K_{ij})$ with gauge fields such that 
$\Sigma_{end}$ is diffeomorphic to ${R}^3$ minus a ball and the
following asymptotic conditions are provided:
\ben
\vert g_{ij}  - \delta_{ij} \vert + r \vert \p_{a}g_{ij} \vert
+ ... + r^k \vert \p_{a_{1}...a_{k}}g_{ij} \vert +
r \vert K_{ij} \vert + ... + r^k \vert \p_{a_{1}...a_{k}}K_{ij} \vert
\le {\cal O}\bigg( {1\over r} \bigg), \\
\vert F_{\alpha \beta} \vert + r \vert \p_{a} F_{\alpha \beta} \vert
+ ... + r^k \vert \p_{a_{1}...a_{k}}F_{\alpha \beta} \vert
\le {\cal O}\bigg( {1 \over r^2} \bigg),\\
\phi + r~\p_a \phi + r^k~\p_{a_1\dots a_k} \phi \le {\cal O}\bigg( {1 \over r^2} \bigg).
\een

We recall that an embedded timelike hypersurface will be called a {\it photon surface} if any null geodesics initially tangent to it, remain
tangent as long as it exists. On the other hand, by the {\it photon sphere} we mean a {\it photon surface} for which the lapse function $N$ is constant on it, as well as,
the auxiliary conditions for the fields emerging in the theories in question are satisfied.
It turns out that an arbitrary spherically symmetric static spacetime admits a photon sphere subject to the condition \cite{cla01}
\be
g_{tt}~\p_r g_{\theta \theta} = g_{\theta \theta}~\p_r g_{tt}.
\ee
Our main aim is to study various features of a photon sphere in the spacetime of the electrically charged dilaton black hole, which line element is given by
\be
ds^2 = - \bigg( 1 -\frac{r_+}{r} \bigg) \bigg(1 - \frac{r_-}{r} \bigg)^{\frac{1 -\alpha^2}{1+ \alpha^2}} dt^2
+ \frac{dr^2}{\bigg( 1 -\frac{r_+}{r} \bigg) \bigg( 1 - \frac{r_-}{r} \bigg)^ {\frac{1 -\alpha^2}{1+ \alpha^2}} } + r\bigg( r - \frac{r_-}{r} \bigg)^{\frac{2 \alpha^2}{1 + \alpha^2}}
 (d\theta^2 + \sin^2 \theta d\varphi^2).
\ee
The dilaton black hole event horizon is located at $r_+ $, while the case of $r_- $ we have another singularity but it can be ignored it
because of the fact that $r_-< r_+$.  On the other hand, the dilaton field is given by the relation
$
e^{2\phi} = e^{-2\phi_0} \bigg(1 - \frac{r_-}{r} \bigg)^{\frac{2\alpha}{1+ \alpha^2}} ,
$
where $\phi_0$ is the dilaton field value as $r \rightarrow \infty$. The mass $M $ and the charge $Q$ are related by the relations
$M = \frac{r_+}{2} + \frac{1-\alpha^2}{1+\alpha^2}~\frac{r_-}{2}$ and
$
Q^2 = \frac{r_+~r_-}{2}~e^{2 \phi_0}.
$
For such a black hole the photon sphere lies outside the black hole event horizon $r_+$, and forms the timelike hypersurface at $r = r_{phs}$ \cite{cla01}
\be
r_{phs} = \frac{1}{4} \bigg[
\bigg( \frac{3 - \alpha^2}{1 +\alpha^2}\bigg) r_- + 3r_ + \sqrt{\bigg( \bigg(\frac{3 - \alpha^2}{1 +\alpha^2}\bigg) r_- + 3r_+\bigg)^2 - \frac{32 r_+r_-}{1+\alpha^2}} 
\bigg].
\ee
It can be seen that for $r_+ > A r_-$, where 
$ A= \frac{2(7 + 3\alpha^2)(1 +\alpha^2)}{(3-\alpha^2)^2 + 9 (1+\alpha^2)}$,
we obtain a single timelike photon sphere.

In such spacetime we define a dilaton-electric static system as a time slice of the static spacetime 
$(R \times M^3, -N^2 dt^2 + g_{ij}dx^i dx^j)$.
Then, one defines the photon surface, the main ingredient of our considerations. Namely, let $(M^3,~g_{ij},~N,~\psi,~\phi)$ be a dilaton-electric system bounded with a static
spacetime defined above, with the line element given by the equation (\ref{lel}). 
Having in mind the aforementioned definition of a {\it photon sphere}, it will be subject to a timelike embedded hypersurface
$(P^3,~h_{ij}) \hookrightarrow (R \times M^3, -N^2 dt^2 + g_{ij}dx^i dx^j)$, if the embedding is umbilic and the lapse function, the electric one-form and dilaton form $d\phi$ are normal to
$P^3$. The {\it photon sphere} emerges as the inner boundary of the spacetime in question \cite{ced15}. Namely
\be
(P^3,~h_{ij}) = (R \times \Sigma^2,~-N^2 dt^2 + \sigma_{ij}dx^i dx^j) = \cup_{i=1}^I~(R \times \Sigma_i^2,~-N_i^2 dt^2 + \sigma^{(i)}_{ij}dx^i dx^j),
\ee
where each $P_i^3$ is a connected component of $P^3$.

In order to conduct the uniqueness proof of the photon sphere in EMD gravity we shall follow the reasoning presented in \cite{ced15}.
In the first step we define dilaton electrostatic system $(M^3,~g_{ij},~N,~\psi,~\phi)$ which will be asymptotic to the dilaton black hole solution and will
have a Killing horizon boundary. It can be performed by gluing pieces of (spatial) dilaton black hole manifold
of the adequate masses, charges and values dilaton field. In order to create a new horizon boundary corresponding to each $\Sigma_i^2$ we attach (glue)
at each photon sphere base $\Sigma_i^2$  a neck piece of dilaton black hole manifold of $\mu_i>0,~Q_i,~\phi$ (the cylindrical piece) between the photon sphere in question and its event horizon.
Away from the gluing surface the manifold will have non-negative scalar curvature, it will be smooth, and the metric lapse function, electric potential 
and dilaton field will also be smooth away from the glued surfaces. In the next step, 
we double the glued manifolds under consideration and assert that the emergent system will be smooth across the boundary.
In the last step we perform the adequate conformal transformations in order to apply the conformal positive energy theorem, which will complete the proof.

\subsection{Asymptotically flat manifold with minimal boundary and non-negative scalar curvature}
We commence with the definition of the Komar type charge in the form provide by
\be
Q_i = - \frac{1}{4 \pi} \int_{\Sigma_i^2} dA~\frac{e^{-\alpha \phi}~n^a~E_{a}^{(i)}}{N} = - \frac{e^{-\alpha \phi}~r_i^2~n^a~E_{a}^{(i)}}{N_i},
\ee
where $n^a$ is a unit normal to $P^3$. The above relation corresponds to the dilaton-elecric charge, while the definition of dilaton charge yields
\be
q_i = - \frac{1}{4 \pi} \int_{\Sigma_i^2} dA~n^j~\na_j \phi^{(i)} = - r_i^2~n^j~\na_j \phi^{(i)} .
\ee
On the other hand, using the equation (36) from Ref.\cite{yaz15bb}, the above definitions enable to find that
\be
\frac{4}{3} = H_i^2~r_i^2 + \frac{4}{3}~\frac{e^{-2 \alpha \phi}}{N_i^2} ~n^j E_k^{(i)} ~n^j E_k^{(i)}
- \frac{4}{3}~r_i^2~n^m {}^{(g)}\na_m \phi^{(i)} ~n^a{}^{(g)}\na_a \phi^{(i)},
\label{ya15}
\ee
where $H_i$ stands for for the mean curvature of each i-th component of $P^3$. The relation (\ref{ya15})
can be rewritten in the form as follows:
\be
\frac{4}{3} = H_i^2~r_i^2  + \frac{4}{3} \bigg( \frac{Q_i}{r_i} \bigg)^2 - \frac{4}{3} \bigg( \frac{q_i}{r_i} \bigg)^2.
\ee
Then, following Ref.\cite{ced15}  we define the mass $\mu_i$ on each $\Sigma_i^2 $and intervals $I_i$
\be
\mu_i = \frac{r_i}{3}, \qquad
I_i = \bigg[ s_i = 2 \mu_i,~r_i = r_{phs}(Q_i,~q_i,~\mu_i) \bigg] \subset R.
\ee
Next, we glue in to each boundary component of $\Sigma_i^2$ a cylinder $I_i \times \Sigma_i^2$. The photon sphere in question
component, i.e., $\Sigma_i^2 \subset M^3$ is related to the level $\{ r_i \} \times \Sigma_i^2 $ of the above constructed cylinder. In what follows we shall
call this surface still $\Sigma_i^2$. By virtue of the aforementioned procedure we obtain the manifold ${\tilde M}^3$ which has the inner boundary 
$$B = \cup_{i=1}^I ~\{ s_i \} \times \Sigma_i^2.$$

In the next step we shall build electro-dilaton system smooth away from the gluing  surface  $\Sigma_i^2$. It should be also geodesically complete up to the corresponding boundary $B$.
Just on the cylinder $I_i \times \Sigma_i^2$ one defines the line element provided by
\be
ds^2 \mid_{I_i \times \Sigma_i^2} = \ga_{ij} dx^i dx^j = \frac{dr^2}{f_i^2(r)} + \frac{g_i(r)}{r_i^2}\sigma_i = \frac{dr^2}{f_i^2(r)} + {g_i(r)}d\Omega^2,
\ee
where we have denoted
\be
f_i(r) = \bigg( 1 -\frac{r_{+(i)}(Q_i,~\mu_i)}{r} \bigg) \bigg( 1 - \frac{r_{-(i)}(Q_i,~\mu_i)}{r} \bigg)^ {\frac{1 -\alpha^2}{2(1+ \alpha^2)}} ,
\qquad
g_i(r) = r\bigg( r - \frac{r_{-(i)}(Q_i,~\mu_i)}{r} \bigg)^{\frac{2 \alpha^2}{1 + \alpha^2}},
\ee
and $\sigma_i = r_i^2~d\Omega$. \\
To conclude, it was glued in the portion of the spatial dilaton black hole system possessing mass $\mu_i >0$ and charge $Q_i$ subject to the non-extremality condition. It was done from the 
gluing surface to the photon sphere in question.\\

The next problem, will be to show that $\ga_{ij} \mid_{I_i \times \Sigma_i^2}$ is smooth away from the glueing surface and it is a function of $C^{1,1}$ class across $\Sigma_i^2$. 
Let us introduce the function \cite{ced15}
\be
\xi : {\tilde M}^3 \rightarrow R: p \rightarrow  \left\{ \begin{array}{ll}
N(p) & \textrm{if  $p \in M^3,$}\\
\frac{3 m_i}{r_i}~f_i(r(p)) & \textrm{if $p \in I_i \times \Sigma_i^2.$}
\end{array} \right.
\ee
One will apply it as a smooth collar function across glueing surfaces. By the construction the function $\xi$ is smooth away from $\Sigma_i^2$, ~for all $i \in \{1,~\dots,~I \}$. The choice of the conformal factor
$\frac{3 m_i}{r_i}$ as well as 
the relation binding i-th mean curvature with $N_i$, i.e.,  $N_i~H_i = 2~n^{(a)} {}^{(g)}\na_a N_i$ (for the derivation of this equation see Refs.\cite{ced15,yaz15b}),
imply that $\xi$ has the same constant value at each of the sides of  $\Sigma_i^2$. Hence it is well defined across $\Sigma_i^2$.

The unit normal to $\Sigma_i^2$ towards the dilaton black hole side has the form
\be
n_r = \xi_i(r_i)~\p_r.
\ee
The definitions of $m_i,~\mu_i$ and charges ensure that the normal derivative of $\xi$ is the same positive constant on the both sides of $\Sigma_i^2$. It means that this fact allows one to implement
the function $\xi$ as smooth coordinate function in the neighborhood of each analyzed $\Sigma_i^2 \subset {\tilde M}^3$.

In order to show that $\xi$ is $C^{1,1}$ class function we shall take into account local coordinates on $\Sigma_i^2$  as well as, a flow to a neighborhood of $\Sigma_i^2 \subset {\tilde M}^3$ along
the level set flow defined by $\xi$. Then, it is enough to show that for all $A,~K = 1,~2$ the components of the metric tensor, $\tq_{AB}, ~\tq_{A \xi},~\tq_{\xi \xi}$, are $C^{1,1}$ class functions, with respect to 
the local coordinates $(x_A,~\xi)$ across the aforementioned $\xi$ function level set of $\Sigma_i^2 $.

Because of the fact that $\p_\xi$ is given by
\be
\p_\xi = \frac{1}{n^a {}^{(\tq)}\na_a \xi} n^j {}^{(\tq)}\na_j,
\ee
the continuity of ${\tq}_{ij}$ in $(x_A,~\xi)$ coordinate system and smoothness in the tangential directions along $\Sigma_i^2 $ is seen. Then, the metric tensor components imply
\be
\tq_{AB} = r_i^2~\Omega_{AB}, \qquad \tq_{A \xi}= 0, \qquad \tq_{\xi \xi} = \frac{1}{(n^a {}^{(\tq)}\na_a \xi)^2},
\ee
on $\Sigma_i^2 $ (from both sides).

Further,  we calculate the derivative of $\tq_{AB}$. It yields
\be
\p_\xi(\tq_{AB}) = \frac{2}{n^a {}^{(\tq)}\na_a \xi}\tth_{AB}.
\ee
In the proceeding sections we show the umbilicity of the every component of any dilaton photon sphere, as well as, the fact that the mean curvature of every 
photon sphere is determined by its radius and charges (up to the signs). Having in mind the exact form of the metric tensor $\tq_{ij}$, one can conclude that 
$\tth = \pm 1/2~H_i \sigma_i =\pm 1/2~H_i~r_i^2~\Omega$, hold on both sides of the dilaton photon sphere. As far as the sign is concerned, 
from the side of $M^3$, ~$H_i > 0$ ($H_i$ is calculated with respect to $n^a$ being directed towards the asymptotic end).
On the dilaton black hole side, the mean curvature of the dilaton photon surface is directed towards infinity and thus into $M^3$. It is also positive . Therefore, in both considered cases $\tth_{AB}$
and thus $\p_\xi(\tq_{AB})$ coincide from the two sides of $\Sigma_i^2 $ \cite{ced15,ced15a}.

The relation $\p_\xi(\tq_{A \xi}) = 0$ holds on both sides of $\Sigma_i^2 $ (by the construction $\tq_{A \xi} =0)$. 
It remains to show that $\p_\xi (\tq_{\xi \xi}) $ coincides on both sides of the hypersurface in question. In order to do so let us calculate
\be
\p_\xi(\tq_{\xi \xi}) = - \frac{2}{(n^a {}^{(\tq)}\na_a  \xi)^5}~(n^j {}^{(\tq)}\na_j)(n^b {}^{(\tq)}\na_b \xi),  
\ee
from both sides of $\Sigma_i^2 $. 

Let us recall that for the isometric embedding
with a unit normal $n_i$ and the second fundamental form $K_{ab}$, for every smooth function $\theta$, we have
\be
(D,~h_{ij},~A^n) \hookrightarrow (\na,~g_{ij},~B^{n+1}), \qquad
\na_a \na^a \theta = D_m D^m \theta + (n^a~\na_a)(n^j~\na_j) \theta + K_{a}{}^a~n^i~\na_i \theta.
\label{ide}
\ee
Using the indentity (\ref{ide}) we obtain
\be
(n^k{}^{(\tq)} \na_k) (n^a{}^{(\tq)} \na_a \xi) = {}^{(\tq)}\na_a {}^{(\tq)}\na^a N - H_i~n^c {}^{(\tq)}\na_c N.
\ee
By virtue of the fact that $\xi$ is constant on $\Sigma_i^2$ and having in mind equation of motion (\ref{eqm1}), we receive
\be
(n^k{}^{(\tq)} \na_k)(n^a{}^{(\tq)} \na_a \xi) = \frac{e^{-2 \alpha \phi}}{ N} {}^{(\tq)}\na_{m}\psi {}^{(\tq)}\na^{m} \psi - H_i~n^c {}^{(\tq)}\na_c N,
\ee
on both sides of $\Sigma_i^2 $. Moreover, we recall that $n^a{}^{(\tq)}\na \psi,~N,~H_i,~n^a{}^{(\tq)}\na_a \xi$ are continuos across $\Sigma_i^2 $,
which in turn implies that $(n^a {}^{(\tq)} \na_a)(n^a{}^{(\tq)} \na_a \xi)$ is continuous across $\Sigma_i^2 $. Consequently, 
 it concludes that $\tq$ is $C^{1,1}$ class  across $\Sigma_i^2 $ and for arbitrary $i \in \{1,~\dots, I \}$ the set $({\tilde M}^3,~\tq_{AB},~N, \psi,~\phi)$ belongs to the same class.

\subsection{Conformal transformations leading to non-negativity of scalar curvature and vanishing of the ADM mass}
In this section we shall consider basic conformal transformations which lead to the conformal positive theorem being the key ingredient in the proof of
the uniqueness of the black hole dilaton {\it photon sphere} \cite{sugra1}. For the brevity of the notation, in this section, we write $\Sigma$ instead of $\Sigma_i^2 $.

To proceed further, let us introduce the definitions of the crucial quantities in the 
the proof of the uniqueness. Namely, they can be written as follows:
\ben
\Phi_{1} &=& \frac{1}{ 2} \bigg[ e^{\alpha \phi} ~N + \frac{1}{e^{\alpha \phi}~N} - (1 + \alpha^2)~\frac{\psi^2}{e^\phi~N} \bigg], \\
\Phi_{0} &=& \sqrt{1 + \alpha^2}~\frac{\psi}{e^{\alpha \phi}~N},\\
\Phi_{-1} &=& \frac{1}{2} \bigg[  e^{\alpha \phi} ~N - \frac{1}{e^{\alpha \phi}~N} - (1 + \alpha^2)~\frac{\psi^2}{e^\phi~N}     \bigg],
\een
and
\ben
\Psi_{1} &=& \frac{1}{2} \bigg[  e^{-\frac{\phi}{\alpha}}~N + \frac{e^{\frac{\phi}{\alpha}}}{N} \bigg],\\
\Psi_{-1} &=& \frac{1}{2} \bigg[  e^{-\frac{\phi}{\alpha}}~N - \frac{e^{\frac{\phi}{\alpha}}}{N} \bigg].
\een
It worth pointing out that defining the metric tensor $\eta_{AB} = diag(1, -1, -1)$, it can be achieved
that $\Phi_{A} \Phi^{A} = \Psi_{A} \Psi^{A} = -1$, where $A = - 1, 0, 1$.
Having in mind the conformal transformation provided by
\be
\tilde g_{ij} = N^{2} g_{ij},
\ee
one can introduce the symmetric tensors written in terms of $\Phi_A$
in the following form:
\be 
\tG_{ij} = \tna_{i} \Phi_{-1} \tna_{j} \Phi_{-1} - \tna_{i} \Phi_{0} \tna_{j} \Phi_{0} - \tna_{i} \Phi_{1} \tna_{j} \Phi_{1},
\label{g1}
\ee
and similarly for the potential $\Psi_{A}$
\be
\tH_{ij} = \tna_{i} \Psi_{-1} \tna_{j} \Psi_{-1} - \tna_{i} \Psi_{1} \tna_{j} \Psi_{1},
\label{h1}
\ee
where by $\tna_{i}$ we have denoted the covariant derivative with respect to the metric $\tg_{ij}$.
Consequently, according to
the relations (\ref{g1}) and (\ref{h1}), the field equations
may be cast in the forms
\be
\tna^{2}\Phi_{A} = \tG_{i}{}{}^{i} \Phi_{A}, \qquad
\tna^{2} \Psi_{A} = \tH_{i}{}{}^{i} \Psi_{A}.
\label{ppff}
\ee
It can be verified by the direct calculations that the Ricci curvature tensor with respect to the conformally rescaled metric $\tg_{ij}$ is given by the relation
\be
\tR_{ij} = \frac{2}{1 + \alpha^2}~\bigg( \tG_{ij} + \alpha^2~\tH_{ij} \bigg).
\label{rr}
\ee
As far as the conformal positive energy theorem is concerned, one assumes that we have to do with two asymptotically flat Riemannian three-dimensional manifolds
$(\Sigma^{\Phi},~ {}^{(\Phi)}g_{ij})$ and $(\Sigma^{\Psi},~ {}^{(\Psi)}g_{ij})$. Moreover, we establish the conformal transformation 
of the form ${}^{(\Psi)}g_{ij} = \Omega^2~{}^{(\Phi)}g_{ij}$, connecting the adequate metric tensors of the manifolds in question. It implies that the corresponding masses
obeys the relation of the form ${}^{\Phi}m + \beta~{}^{\Psi}m \geq 0$ if ${}^{(\Phi)} R + \beta~\Omega^2~{}^{(\Psi)} R \geq 0$, for some
positive constant $\beta$. The aforementioned inequalities are satisfied it the 
three-dimensional Riemannian manifolds are flat \cite{sim99}. \\
To proceed further, due to the requirement of the conformal positive energy theorem, we introduce
conformal transformations fulfilling the following:
\be
{}^{(\Phi)}g_{ij}^{\pm} = {}^{(\Phi)}\omega_{\pm}^{2}~ \tg_{ij},
\qquad
{}^{(\Psi)}g_{ij}^{\pm} = {}^{(\Psi)}\omega_{\pm}^{2}~ \tg_{ij}.
\ee
Their conformal factors are subject to the relations of the forms
\be
{}^{(\Phi)}\omega_{\pm} = {\Phi_{1} \pm 1 \over 2}, \qquad
{}^{(\Psi)}\omega_{\pm} = {\Psi_{1} \pm 1 \over 2}.
\label{pf}
\ee
Next, we implement the standard procedure of pasting $(\Sigma_{\pm}^{\Phi},~ {}^{(\Phi)}g_{ij}^{\pm})$ and 
$(\Sigma_{\pm}^{\Psi},~ {}^{(\Psi)}g_{ij}^{\pm})$ across their shared minimal boundary \cite{ced15}. 
We have four manifolds $(\Sigma_{+}^{\Phi},~ {}^{(\Phi)}g_{ij}^{+})$,
$(\Sigma_{-}^{\Phi},~ {}^{(\Phi)}g_{ij}^{-})$,~ $(\Sigma_{+}^{\Psi},~ {}^{(\Psi)}g_{ij}^{+})$,~$(\Sigma_{-}^{\Psi},~ {}^{(\Psi)}g_{ij}^{+})$.
Pasting them across shared minimal boundaries ${\cal B}^\Psi$ and ${\cal B}^\Phi$, one can construct complete regular hypersurfaces
$\Sigma^{\Phi} = \Sigma_{+}^{\Phi} \cup \Sigma_{-}^{\Phi}$ and $\Sigma^{\Psi} = \Sigma_{+}^{\Psi} \cup \Sigma_{-}^{\Psi} $. 
Having two regular hypersurfaces one has to check that each total gravitational mass on $\Sigma^{\Phi}$ and on $\Sigma^{\Psi}$ vanishes.

In order to find this result we shall use the conformal positive theorem \cite{sim99}. 
 On this account, it is customary to define 
another conformal transformation described by the relation
\be
{\hat g}^{\pm}_{ij} = \bigg[ \bigg( {}^{(\Phi)}\omega_{\pm} \bigg)^2
 \bigg( {}^{(\Psi)}\omega_{\pm} \bigg)^{2 \alpha^2} \bigg]^{1 \over 2}\tg_{ij},
\ee
it follows that 
the Ricci curvature tensor on the space under consideration can be written in the form as
\ben \label{ric}
(1+\alpha^2)~\hat R_\pm &=& \bigg[ {}^{(\Phi)}\omega_{\pm}^2~ {}^{(\Psi)}\omega_{\pm}^{2 \alpha^2} \bigg]
^{-{1 \over 2}}
\bigg( {}^{(\Phi)}\omega_{\pm}^{2} {}^{(\Phi)}R_\pm +
{}^{(\Psi)}\omega_{\pm}^{2} {}^{(\Psi)}R_\pm \bigg) \\ \nonumber
&+& \frac{2 \alpha^2}{1+\alpha^2}~
\bigg( \hat \na _{i} \ln {}^{(\Phi)}\omega_{\pm} - {\hat \na} _{i} \ln {}^{(\Psi)}\omega_{\pm} \bigg)  
\bigg( \hat \na ^{i} \ln {}^{(\Phi)}\omega_{\pm} - {\hat \na}^{i} \ln {}^{(\Psi)}\omega_{\pm} \bigg).  
\een
Further, by the direct calculations it reveals that  the equation (\ref{ric}) can be cast as follows:
\ben
{}^{(\Phi)}\omega_{\pm}^{2}~ {}^{(\Phi)}R_\pm + \alpha^2~{}^{(\Psi)}\omega_{\pm}^{2}~ {}^{(\Psi)}R_\pm &=& 
2~\mid {\Phi_{0} \tna_{i} \Phi_{-1}
- \Phi_{-1} \tna_{i} \Phi_{0} \over
\Phi_{1} \pm 1 } \mid^2 \\ \nonumber
&+&
2~\mid { \Psi_{0} \tna_{i} \Psi_{-1}
- \Psi_{-1} \tna_{i} \Psi_{0} \over
\Psi_{1} \pm 1} \mid^2,
\een
that the terms on the right-hand side of the relation are non-negative.

On the other hand, the conformal positive energy theorem 
enables us to claim that ${}^{(\Phi)}\omega = const.~{}^{(\Psi)}\omega$, as well as, $\Phi_{0} = const~ \Phi_{-1}$ and $\Psi_{0} = const~ \Psi_{-1}$.
Moreover, each of the manifolds $(\Sigma^{\Phi},~ {}^{\Phi}g_{ij})$, $(\Sigma^{\Psi},~ {}^{\Psi}g_{ij})$ and
$(\hat \Sigma,~ {\hat g}_{ij})$ are flat.
Just, the manifold $(\Sigma,~ g_{ij})$ is conformally flat.
The metric tensor $\hat g_{ij}$ can be written in  conformally flat form. Namely, let us define
\be
\hat g_{ij} = {\cal U}^{4}~ {}^{(\Phi)}g_{ij},
\label{gg}
\ee
where one sets ${\cal U} = ({}^{\Phi}\omega_{\pm} V)^{-1/2}$.
The fact that the Ricci scalar in $\hat g_{ij}$ metric is equal to zero implies that the equations of motion of the system in question reduce
to the Laplace equation on the three-dimensional Euclidean manifold
\be
\na_{i}\na^{i}{\cal U} = 0,
\ee
where $\na$ is the connection on a flat manifold. 
Next, it yields that the expression for the flat base space is valid, i.e., one obtains the following:
\be
{}^{(\Phi)}g_{ij} dx^{i}dx^{j} = \trho^{2} d{\cal U}^2 + {\tilde h}_{AB}dx^{A}dx^{B},
\ee
The photon sphere will be located at some constant value of ${\cal U}$.
The radius of the photon sphere can be given at the fix value of
$\rho$-coordinate \cite{ced14}.
All these enable that on the hypersurface $\Sigma$ the metric tensor can be given in the form of 
$$\hat g_{ij}dx^{i}dx^{j} = \rho^2 dV^2 + h_{AB}dx^{A}dx^{B},$$
and a connected component of the photon surface can be identify at fixed value of $\rho$-coordinate.\\
In order to proceed further, let us assume that
${\cal U}_{1}$ and ${\cal U}_{2}$ consist
two solutions of the boundary value problem of the system in question.
Using Green identity and integrating over the volume element, we arrive at the relation
\be
\bigg( \int_{r \rightarrow \infty} - \int_{\cal H} \bigg) 
\bigg( {\cal U}_{1} - {\cal U}_{2} \bigg) {\p \over \p r}
\bigg( {\cal U}_{1} - {\cal U}_{2} \bigg) dS = \int_{\Omega}
\mid \na \bigg( {\cal U}_{1} - {\cal U}_{2} \bigg) \mid^{2} d\Omega.
\ee
In view of the last equation, the surface integrals disappear due to the imposed boundary conditions.
On the other hand, by virtue of the above relation one finds that
the volume integral must be identically equal to zero. Taking all the above into account, we can assert that the following theorem holds: \\ 
{\bf Theorem}:\\
Let us consider the set $(M^3,~g_{ij},~N,~\psi, ~\phi)$ being the system asymptotic to the dilaton black hole spacetime and possessing the photon sphere
$(P^3,~h_{ij}) \hookrightarrow (R \times M^3,~ -N^2 dt^2 + g_{ij}dx^i dx^j)$, which can be regarded as the inner boundary of $R \times M^3$. 
Suppose further that $M$ and $Q$ are the ADM mass and the total charge of $(R \times M^3,~ -N^2 dt^2 + g_{ij}dx^i dx^j)$.
Then, $(R \times M^3,~ -N^2 dt^2 + g_{ij}dx^i dx^j)$ is isometric to the region of 
$r \ge r_{phs}, $ 
exterior to the {\it photon sphere} 
in the electrically charged dilaton black hole spacetime. The photon sphere in question is connected and it constitutes a cylinder over a topological sphere.

\section{Conclusions}
In our paper we have elaborated the uniqueness of a static asymptotically flat black hole photon sphere in Einstein-Maxwell-dilaton theory of gravity with arbitrary coupling constant
$\alpha$. Using the conformal positive energy theorem,
we show that the region exterior to the photon sphere of the adequate radius in the electrically charged dilaton black hole spacetime is connected and it
authorizes a cylinder over a topological sphere. The proof is valid for non-extremal dilaton black hole photon sphere.




\begin{acknowledgments}
MR was partially supported by the grant of the National Science Center $DEC-2014/15/B/ST2/00089$.
\end{acknowledgments}



\end{document}